\documentclass[12pt,a4paper,oneside]{article}  
\usepackage[english]{babel}
\usepackage{amsmath,amsfonts}
\newcommand{\vett}[1]{\mathbf{#1}}
\newcommand{\im}{i}

\newcommand{\Frac}[2]{\displaystyle\frac{#1}{#2}}

\title{\vspace{-3\baselineskip}%
On the pairing structure of the vacuum induced by a magnetic field 
in $2+1$-dimensional Dirac field theory} 

\author{Giovanni Jona-Lasinio\footnote{\ 
Email: {\tt Jona@roma1.infn.it}} \\
Dipartimento di Fisica, Universit\`a di Roma ``La Sapienza''\\ 
P.le Aldo Moro, 2 --- 00185 Roma  
\and
Francesca Maria Marchetti\footnote{\ 
Email: {\tt fmarchet@cibs.sns.it}} \\
Scuola Normale Superiore \\ 
P.zza dei Cavalieri, 7 --- 56126 Pisa} 

\date{\small \emph{Roma, \today}}

\begin{document}

\maketitle

%\tableofcontents

\begin{abstract} 
\emph{Using a sequence of Bogoliubov transformations, we obtain an 
exact expression for the vacuum state of a Dirac field in $2+1$ 
dimensions in presence of a constant magnetic field. This 
expression reveals a peculiar two level pairing structure for any 
value of the mass $m \ge 0$. This calculation clarifies the nature 
of the condensate in the lowest Landau level whose existence has 
been emphasized recently by several authors}.
\end{abstract} 

\section{Introduction} 
\label{intro} 

The phenomenon of symmetry breaking induced by an external  
homogeneous magnetic field of an internal $U(2)$ symmetry for the  
Dirac field in $2+1$ dimensions has recently attracted some  
attention. The reason why people are interested in this phenomenon  
is its possible relevance in the more complex context of  
interacting fields, especially in situations where spontaneous  
symmetry breaking is expected, like in strong coupling QED and NJL  
models 
\cite{mira1,kriv1,parw1,dunne}.The 
symmetry breaking is connected with a property of the theory 
called spectral asymmetry of the lowest Landau level (LLL) and it 
is revealed by the calculation of the order parameter  
\begin{equation} 
\lim_{m \to 0} \langle B| \bar{\psi}^B (x) \psi^B (x) |B \rangle = 
- \Frac{|eB|}{2 \pi} 
\label{prima}
\end{equation}
where $m$ is the mass of the Dirac field. This parameter is 
sometimes called \emph{flavor condensate}.  

The expression \eqref{prima} can be obtained using the proper time 
method of Schwinger \cite{schwi} or through the development the 
propagator in terms of Landau poles \cite{chodo,mira1}. From the 
calculations one sees that the result \eqref{prima} depends on the 
dominance of the LLL, when $m \to 0$. 

The methods used to obtain \eqref{prima} do not reveal completely 
the structure of the condensate since they do not need the 
explicit expression of the vacuum state. In this paper we obtain 
the explicit expression of the vacuum in presence of a magnetic 
field in terms of the creation and destruction operator of the 
free field ($B=0$). The problem is not entirely trivial and is 
solved by performing in sequence two Bogoliubov transformations. 
The first leads from the vacuum of the free field $|0 \rangle$ to 
a BCS type state $\widehat{|0 \rangle}$, pairing particles and 
antiparticles with opposite momenta. This step does not involve 
the magnetic field. The second transformation acts on auxiliary 
creation and destruction operators associated with $\widehat{|0 
\rangle}$. Their pairing leads to the vacuum $|B \rangle$ of the 
Dirac field in presence of magnetic field. These operators when 
applied to $\widehat{|0 \rangle}$ create or destroy states having 
the same quantum numbers of the Landau levels; they can be 
interpreted as degenerate eigenstates of an auxiliary hamiltonian 
$\widehat{\mathrm{H}}$ with energy equal to the LLL. The structure 
of the vacuum just described is the same for any value of $m$ 
including ${m \to 0}$ and disappears in the non relativistic limit 
that is for ${m \to \infty}$. 

We believe that this calculation has a twofold  
interest. The problem of the Dirac field in $2+1$ dimensions in 
presence of a constant magnetic field is solved by operator 
techniques which make clear its non perturbative nature. 
In view of the study of non trivial models with 
interaction it provides an application of the principle 
\emph{solve the ground state first, then it is easier to 
understand the excitations} \cite{shury}.   

\section{The Dirac equation in $2+1$ dimensions}
\label{notaz}

The main properties of the $2+1$ dimensional Dirac field theory in 
an external homogeneous magnetic field have been discussed in 
\cite{bineg,boyan,appel,mira1} and will be now summarized to make 
the paper self-contained.  

Consider the lagrangian $\mathcal{L}$: 
\begin{equation} 
\begin{split} 
\mathcal{L}     &= \bar{\psi}^B (x) \left[ \im \gamma^\mu 
\mathcal{D}_\mu - m  \right] \psi^B (x) \\ 
\mathcal{D}_\mu &= \partial_\mu - \im e A_\mu 
\end{split}
\label{elle1}
\end{equation}
where $e$ is the modulus of the electron charge. Since the 
external magnetic field is constant and homogeneous we can choose 
the Landau gauge:   
\begin{equation} 
\label{gaug1} A_\mu = -\delta_{\mu 1} Bx^2 \ .
\end{equation}

In $2+1$ dimensions there are two inequivalent \emph{minimal 
versions} of the Dirac algebra \cite{appel}:  
\begin{equation}
\begin{array}{cl}
\{ \gamma^\mu , \gamma^\nu \} = 2 g^{\mu \nu} & \mu  = 0,1,2 \\ 
{\gamma^0}^\dag = \gamma^0  \qquad {\gamma^i}^\dag = - \gamma^i  & 
i  = 1,2 \ . 
\end{array}
\label{cliff}
\end{equation}
In the following we shall use the representations:
\begin{gather} 
{\widetilde{\gamma}}^0 = \sigma_3 \qquad {\widetilde{\gamma}}^1 =  
\im \sigma_1 \qquad {\widetilde{\gamma}}^2 =  \im \sigma_2 
\label{comp1}\\
\intertext{and}
{\widehat{\gamma}}^0 = -\sigma_3 \qquad {\widehat{\gamma}}^1 = - 
\im \sigma_1 \qquad {\widehat{\gamma}}^2 = - \im \sigma_2 
\label{comp2} 
\end{gather}
where  $\sigma_i$ are the Pauli matrices. The complex spinors 
associated with the two minimal versions will be respectively 
indicated with $\psi_1 (x)$ e $\psi_2 (x)$. 

A \emph{chiral version} is obtained as the direct sum of the two 
inequivalent irreducible representations \eqref{comp1} and 
\eqref{comp2}:  
\begin{equation}
\gamma^\mu = \begin{pmatrix} {\widetilde{\gamma}}^\mu & 0 \\ 0 & 
{\widehat{\gamma}}^\mu \end{pmatrix} = \begin{pmatrix} 
{\widetilde{\gamma}}^\mu & 0 \\ 0 & -{\widetilde{\gamma}}^\mu 
\end{pmatrix}  \qquad \psi(x) = \begin{pmatrix} \psi_1(x) \\ 
\psi_2(x) \end{pmatrix} \ . 
\label{comp3}
\end{equation}

With this notation the lagrangian in the chiral version is 
connected to the lagrangians in the two minimal versions by the 
following relationship:  
\begin{equation}
\mathcal{L} = \mathcal{L}_1 + \mathcal{L}_2  \qquad \mathcal{L}_i 
= \bar{\psi}^B_i (x) \left[ \im \widetilde{\gamma}^\mu 
\mathcal{D}_\mu - \alpha_i m \right] \psi^B_i (x) 
\label{lagr1}
\end{equation}
\begin{equation}
\alpha_i = \begin{cases} 1 & i=1 \\ -1 & i=2 \end{cases} \qquad 
\bar{\psi}^B_i(x) = {\psi^B_i}^\dag (x) \sigma_3 \qquad \forall 
i=1,2 \ . 
\end{equation}
The two lagrangians $\mathcal{L}_1$ and $\mathcal{L}_2$ differ 
only for the sign of the mass term.  

In the minimal versions both lagrangians $\mathcal{L}_1$ and  
$\mathcal{L}_2$ are invariant under charge conjugation 
$\mathcal{C}$, while the mass term changes sign under parity 
$\mathcal{P}$ and time reversal $\mathcal{T}$. In the minimal 
versions there is no generator of chiral symmetry, that is there 
is no symmetry (except $\mathcal{P}$ and $\mathcal{T}$), that 
distinguishes the theory with mass from the theory with zero mass.  

In the chiral version the three discrete symmetries $\mathcal{C}$, 
$\mathcal{P}$ and $\mathcal{T}$ are preserved and therefore the 
mass term breaks neither parity nor time reversal. Besides, we can 
introduce the following matrices, which anticommute with each 
other and with the other matrices $\gamma^\mu$ defined in 
\eqref{comp3}:  
\begin{equation}
\gamma^3 = \im \begin{pmatrix} 0 & \mathbb{I} \\ \mathbb{I} & 0 
\end{pmatrix} \qquad \gamma^5 = \im \gamma^0 \gamma^1 \gamma^2 
\gamma^3 = \im \begin{pmatrix} 0 & \mathbb{I} \\ - \mathbb{I} & 0 
\end{pmatrix} \ . 
\end{equation}  

In the chiral version we can introduce an internal $U(2)$ group of 
symmetry, defined by the generators $T_\alpha$: 
\begin{equation}
U(\vett{\omega}) \in U(2) : \ U(\vett{\omega}) = e^{\im 
\omega^\alpha T_\alpha} \qquad \alpha = 0,...,3 
\label{flavo}
\end{equation}
\begin{equation}
T_0 = \mathbb{I}  \qquad  T_1 = \gamma^5  \qquad T_2 = 
\Frac{1}{\im} \gamma^3 \qquad  T_3 = \gamma^3 \gamma^5 \ . 
\label{flav1}
\end{equation}
The mass term $m\bar{\psi}\psi$ of the lagrangian breaks this 
symmetry group into the subgroup $U(1) \times U(1)$ with 
generators $T_0$ and $T_3$.  
In the limit where the mass $m$ tends to zero, the $U(2)$ symmetry 
is not recovered and one obtains a continuum of degenerate vacua. 
The result \eqref{prima} can be interpreted as an effect of spin 
polarization (or spin orientation) due to the magnetic field. A 
generalization of formula \eqref{prima} in which the magnetic 
flux $\int d^2 x B(\vett{x})$ replaces $B$ is valid in the 
inhomogeneous case \cite{parw1,dunne}. 

The problem of a free Dirac field, minimally coupled to a 
homogeneous magnetic field, can be exactly solved and in the 
Landau gauge the expression of the Dirac field in the chiral 
version is ($eB>0$): 
\begin{gather}
\label{psim1}
\psi^B (x) = \begin{pmatrix} \psi_1^B (x) \\ \psi_2^B (x) 
\end{pmatrix} \\ 
\psi_1^B (x) = \sum_{n=0}^{\infty} \sum_{p^1} \{u_{np^1}(x) 
a_{np^1} + v_{n-p^1}(x) b_{np^1}^\dag \} \\ 
\psi_2^B (x) = \sum_{n=0}^{\infty} \sum_{p^1} \{ u_{n p^1}^{(2)} 
(x) c_{np^1} + v_{n -p^1}^{(2)} (x) d_{np^1}^\dag \} 
\label{psim2}
\end{gather}
\begin{equation}
\begin{split}
u_{np^1}(x) &= \Frac{1}{\sqrt{lL_1}}e^{-\im E_nt}e^{\im p^1 x^1} 
\begin{pmatrix} A_n w_n(\xi_{x^2}^{p^1}) \\ -\im B_n w_{n-
1}(\xi_{x^2}^{p^1}) \end{pmatrix} \\ 
v_{np^1}(x) &= \Frac{1}{\sqrt{lL_1}}e^{+\im E_nt}e^{\im p^1x^1} 
\begin{pmatrix} B_n w_n(\xi_{x^2}^{p^1}) \\ +\im A_n w_{n-
1}(\xi_{x^2}^{p^1}) \end{pmatrix} 
\end{split}
\label{enepm}
\end{equation}
\begin{equation}
u_{np^1}^{(2)} (x) = (-1)^n v_{n-p^1} (-x) \quad v_{np^1}^{(2)} 
(x) =(-1)^n u_{n-p^1} (-x) 
\end{equation}
\begin{equation}
A_n = \sqrt{\Frac{E_n + m}{2 E_n}} \qquad B_n = \sqrt{\Frac{E_n - 
m}{2 E_n}} 
\label{defin}
\end{equation}
\begin{equation}
E_n = \sqrt{m^2 + 2neB} \qquad \xi_{x^2}^{p^1} = \Frac{x^2}{l} + 
lp^1 = \sqrt{eB} x^2 + \Frac{p^1}{\sqrt{eB}} 
\end{equation}
\begin{equation}
w_n (\xi) = c_n e^{-\xi^2 / 2} H_n (\xi) = \left( 2^n n! 
\sqrt{\pi} \right)^{-1/2} e^{-\xi^2 / 2} H_n (\xi) \ . 
\end{equation}
The operators $a_{n p^1} , ... , d_{np^1}$ satisfy the usual 
canonical anticommutation relations and $H_n (\xi)$ are the 
Hermite polynomials. The case $eB<0$ can be obtained by applying 
the charge conjugation operator.

One of the main properties of the theory in the minimal versions 
is the spectrum asymmetry concerning the lowest Landau level 
(LLL). This property holds also for an inhomogeneous magnetic 
field \cite{aharo,jacki,thall}. For example, if $m>0$ (that is in 
the representation \eqref{comp1}), and if $eB>0$, the energy 
spectrum is described by: 
\begin{equation*} 
E = \pm \sqrt{m^2 + (2n + 1 - \sigma)eB} \qquad \begin{cases} 
\sigma = +1 \qquad \text{if} \ E>0 \\ \sigma = -1 \qquad \text{if} 
\ E<0  
\end{cases}
\end{equation*}
where $(1/2)\sigma$ are the eigenvalues of the spin operator $S_3 
= (1/2)\sigma_3$. From \eqref{enepm} $u_{0p^1} \ne 0$ and 
$v_{0p^1} \equiv 0$. The situation is inverted in the 
representation \eqref{comp2} ($m < 0$).

\section{Vacuum state}
\label{vuoto}
The lagrangian, describing the Dirac field in presence of an 
external magnetic field, is bilinear in the fermionic field 
$\psi^B (x)$, and we may expect that the vacuum state of the 
theory can be calculated by an approach `` \`a la Bogoliubov'', that 
is by relating the fermionic field in the presence of a magnetic 
field to the free field by means of a linear transformation. The 
calculation, which we will now illustrate, will refer to the 
minimal version representation \eqref{comp1}, as the 
generalization to the chiral version is immediate. 

The Dirac equations describing in $2+1$ dimensions the free field 
and the field in presence of the external field are first order 
differential equations, and it is possible to impose the same 
initial condition: 
\begin{equation}
\psi_1 (0 , \vett{x}) = \psi_1^B (0 , \vett{x}) \ .
\label{conin}
\end{equation}
For the free field we have the plane waves decomposition: 
\begin{equation}
\psi_1 (x) = \sum_{\vett{p}} \sqrt{\Frac{m}{L_1L_2 E_{\vett{p}}}} 
\left\{ u(\vett{p}) e^{-\im p \cdot x}a_{\vett{p}} + v(\vett{p}) 
e^{\im p \cdot x} b_{\vett{p}}^\dag \right\} 
\label{svicm}
\end{equation}
\begin{equation}
u (\vett{p}) = \displaystyle \sqrt{\Frac{E_\vett{p} + 
m}{2m}}\begin{pmatrix}1 \\ \Frac{p^2-\im p^1}{E_\vett{p} + m} 
\end{pmatrix}  \quad v (\vett{p}) = \displaystyle 
\sqrt{\Frac{E_\vett{p} + m}{2m}}\begin{pmatrix} \Frac{p^2+\im 
p^1}{E_\vett{p} + m} \\ 1 \end{pmatrix} 
\label{ugiug}
\end{equation}
where $E_{\vett{p}} = (m^2 + |\vett{p}|^2)^{1/2}$. 

The initial condition \eqref{conin} enables us to find the 
canonical transformation between the operators $(a_{np^1} , 
b_{np^1})$ and the corresponding creation and destruction 
operators in absence of the magnetic field $(a_{\vett{p}} , 
b_{\vett{p}})$. The free field $\psi_1 (x)$ and the one in 
presence of a magnetic field $\psi_1^B (x)$ are not developed in 
the same bases of wave functions, and to obtain the desired 
canonical transformation the following relations are useful: 
\begin{gather}
\label{trfo1}
e^{\im p^2 x^2} = e^{-\im l^2 p^1 p^2} \sqrt{2\pi} 
\sum_{n=0}^\infty (\im)^n w_n ( \xi_{x^2}^{p^1} ) w_n (lp^2) \\ 
w_n (\xi_{x^2}^{p^1}) = \frac{\sqrt{2\pi} (-\im)^n l}{L_2} 
\sum_{p^2} w_n (lp^2) e^{\im p^2 x^2} e^{\im l^2 p^1 p^2} \ . 
\label{trfo2}
\end{gather} 

The expression \eqref{trfo1} follows by analytic continuation for 
$t \to i$ of the relation \cite{hermi}: 
\begin{equation}
\sqrt{\pi} \sum_{n=0}^\infty t^n w_n (x) w_n (y) = 
\Frac{1}{\sqrt{1 - t^2}} e^{\frac{x^2 - y^2}{2} - \frac{(x - 
yt)^2}{1 - t^2}} \quad |t|<1 
\end{equation}
whereas \eqref{trfo2} represents the Fourier transformation of a 
Hermite function. 

Using \eqref{trfo1} and identifying the coefficients associated to 
any function $w_n (\xi)$, we obtain  the transformation
\begin{equation}
\left\{
\begin{split}
a_{np^1} &= \displaystyle \sum_{p^2} \Frac{(\im)^n 
D_{\vett{p}}}{\sqrt{L_2}} \left\{ \alpha_{n \vett{p}} a_{\vett{p}} 
+ \beta_{n \vett{p}} b_{- \vett{p}}^\dag \right\} \\  
b_{n -p^1}^\dag &= \displaystyle \sum_{p^2} \Frac{(\im)^n 
D_{\vett{p}}}{\sqrt{L_2}} \left\{ \delta_{n -\vett{p}}^\ast 
a_{\vett{p}} + \gamma_{n -\vett{p}}^\ast b_{- \vett{p}}^\dag 
\right\}  
\end{split}
\right.
\label{trbo2}
\end{equation}
\begin{equation}
\text{where} \quad 
\begin{split} 
\alpha_{n \vett{p}} &= A_n w_n (lp^2) + B_n C_{\vett{p}}^\ast 
w_{n-1} (lp^2) \\ 
\beta_{n \vett{p}}  &= -A_n C_{\vett{p}} w_n (lp^2) + B_n w_{n-1} 
(lp^2) \\ 
\gamma_{n \vett{p}} &= B_n C_{\vett{p}}^\ast w_n (-lp^2) - A_n 
w_{n - 1} (- lp^2) \\ 
\delta_{n \vett{p}} &= B_n w_n (-lp^2) + A_n C_{\vett{p}} w_{n-1} 
(- lp^2)  
\end{split}
\label{dabgd}
\end{equation}
\begin{equation}
C_{\vett{p}} = \Frac{p^2 + \im p^1}{E_{\vett{p}} + m} \qquad 
D_{\vett{p}} = \sqrt{2\pi l} e^{-\im l^2 p^1 p^2} 
\sqrt{\Frac{E_{\vett{p}} + m}{2 E_{\vett{p}}}} \ . 
\label{defcd}
\end{equation}
These transformations preserve the canonical anticommutation 
relations.

We look for a normalized state $|B \rangle$ such that:
\begin{equation}
a_{np^1} |B \rangle = 0 = b_{np^1} |B \rangle \qquad \forall n,p^1  
\label{cvucm}
\end{equation}

Introducing the auxiliary creation and destruction operators   
\begin{equation} 
\begin{split} 
\widehat{a}_{np^1}  &\equiv \sum_{p^2} \Frac{(\im)^n 
D_{\vett{p}}}{\sqrt{L_2}} w_n (lp^2) \left\{ a_{\vett{p}} - 
C_{\vett{p}} b_{-\vett{p}}^\dag \right\} \\ 
\widehat{b}_{n-p^1} &\equiv \sum_{p^2} \Frac{- (\im)^n 
D_{\vett{p}}^\ast}{\sqrt{L_2}} w_{n-1} (lp^2) \left\{ C_{\vett{p}} 
a_{\vett{p}}^\dag + b_{-\vett{p}} \right\}   
\end{split}
\label{defos}
\end{equation}
the transformation \eqref{trbo2} can be written:
\begin{equation}
\left\{
\begin{split}
a_{np^1} &= A_n \widehat{a}_{np^1} - B_n \widehat{b}_{n -p^1}^\dag 
\\ 
b_{n -p^1} &= A_n \widehat{b}_{n -p^1} + B_n \widehat{a}_{n 
p^1}^\dag \ . 
\end{split}
\right.
\label{refa1}
\end{equation}
with $A_n$ and $B_n$ given in \eqref{defin}. If $\widehat{|0 
\rangle}$ is the vacuum associated with the auxiliary operators  
\begin{equation}
\widehat{a}_{np^1} \widehat{|0 \rangle} = 0 = \widehat{b}_{n-p^1} 
\widehat{|0 \rangle} \qquad \forall n,p^1  
\label{devos}
\end{equation}
we deduce immediately the relation between $|B \rangle$ and 
$\widehat{|0 \rangle}$: 
\begin{equation}
|B \rangle = \prod_{n \ge 1} \prod_{p^1} \left( A_n + B_n 
{\widehat{a}_{n p^1}}^\dag {\widehat{b}_{n-p^1}}^\dag \right) 
\widehat{|0 \rangle} \ . 
\label{revbs}
\end{equation}

The last step consists in finding the relation between the vacuum 
$\widehat{|0 \rangle}$ and the vacuum $|0 \rangle$ in absence of 
magnetic field. We find easily using \eqref{defos} 
\begin{equation}
\widehat{|0 \rangle} = \prod_{\vett{p}} \sqrt{\Frac{E_{\vett{p}} + 
m}{2 E_{\vett{p}}}} \left( 1 + C_{\vett{p}} a_{\vett{p}}^\dag b_{-
\vett{p}}^\dag \right) |0 \rangle \ . 
\label{vucs1}
\end{equation}

The final result is:
\begin{equation}
|B \rangle = \prod_{n \ge 1 \  p^1} \left( A_n + B_n {\widehat{a}_{n 
p^1}}^\dag {\widehat{b}_{n-p^1}}^\dag \right) \prod_{\vett{h}} 
\sqrt{\Frac{E_{\vett{h}} + m}{2 E_{\vett{h}}}} \left( 1 + 
C_{\vett{h}} a_{\vett{h}}^\dag b_{-\vett{h}}^\dag \right) |0 
\rangle \ . 
\label{vucs2}
\end{equation}
The independence of $\widehat{| 0 \rangle}$ on the magnetic field 
is a consequence of the fact that in \eqref{defos} $B$ appears 
only in the multiplicative factors of the Bogoliubov 
transformations within curly brackets. It is clear that the 
structure of the vacuum does not change when $m \to 0$.

In order to clarify the meaning of the auxiliary operators let us 
write the hamiltonian in presence of magnetic field in terms of 
these operators. Using \eqref{defos} we find 
\begin{equation}
\begin{split}
\displaystyle \mathrm{H}^B &= \widehat{\mathrm{H}} - 
\sum_{n=1}^\infty \sum_{p^1} \sqrt{2neB} \left( 
{\widehat{a}_{np^1}}^\dag {\widehat{b}_{n-p^1}}^\dag + 
\widehat{b}_{n-p^1} \widehat{a}_{np^1} \right) \\ 
\displaystyle \widehat{\mathrm{H}} &\equiv m \sum_{n=0}^\infty 
\sum_{p^1} \left( {\widehat{a}_{np^1}}^\dag \widehat{a}_{np^1} + 
{\widehat{b}_{np^1}}^\dag \widehat{b}_{np^1} \right) \ . 
\end{split}
\label{mamma}
\end{equation}

The expression of the hamiltonian $\widehat{\mathrm{H}}$ shows that 
$\widehat{a}_{np^1}$ and $\widehat{b}_{np^1}$ describe particles 
with degenerate energy $m$, that is the energy of the LLL. The 
magnetic field induces creation and destruction of 
particle-antiparticle pairs of the auxiliary field and this 
removes the degeneracy giving the usual Landau levels. 

The auxiliary field $\widehat{\psi} (x)$ is  
\begin{equation} 
\widehat{\psi} (x) \equiv \sum_{n p^1} \left\{ \widehat{u}_{np^1} 
(x) \widehat{a}_{np^1} + \widehat{v}_{n-p^1} (x) 
\widehat{b}_{np^1}^\dag \right\} 
\end{equation}
with
\begin{equation}
\begin{split}
\widehat{u}_{np^1} (x) &= \Frac{1}{\sqrt{l L_1}} e^{- \im m t} 
e^{\im p^1 x^1} w_n (\xi_{x^2}^{p^1}) \begin{pmatrix} 1 \\ 0  
\end{pmatrix} \\ 
\widehat{v}_{np^1} (x) &= \Frac{\im}{\sqrt{l L_1}} e^{\im m t} 
e^{\im p^1 x^1} w_{n-1} (\xi_{x^2}^{p^1}) \begin{pmatrix} 0 \\ 1 
\end{pmatrix}  
\end{split}
\end{equation}

Finally the expression of the vacuum in the chiral version is 
\begin{multline} 
|B \rangle = \prod_{n \ge 1} \prod_{p^1} \left( A_n + B_n 
{\widehat{a}_{np^1}}^\dag {\widehat{b}_{n-p^1}}^\dag \right) 
\left( A_n - B_n {\widehat{c}_{np^1}}^\dag {\widehat{d}_{n-
p^1}}^\dag \right) \cdot \\ 
\cdot \prod_{\vett{h}} \Frac{E_{\vett{h}} + m}{2 E_{\vett{h}}} 
\left( 1 + C_{\vett{h}} a_{\vett{h}}^\dag b_{-\vett{h}}^\dag 
\right) \left( 1 + C_{\vett{h}}^\ast c_{\vett{h}}^\dag d_{-
\vett{h}}^\dag \right) |0 \rangle \ . 
\end{multline}
When $m \to \infty$, $A_n \to 1$, $B_n \to 0$, $C_{\vett{h}} 
\to 0$ and $|B \rangle \to |0 \rangle$.

\end{document}